\documentclass[prb,showpacs,floatfix]{revtex4-1}

\usepackage{graphicx}

\usepackage{graphicx}
\usepackage{color}

\begin{document}

\title{Two Dimensional Honeycomb Materials: random fields, dissipation and fluctuations}

\author{T. Frederico$^1$, O. Oliveira$^{2,1}$, W. de Paula$^1$,  M. S. Hussein$^{1,3}$, T. R. Cardoso$^4$}
\affiliation{$^1$Dep. de F\'\i sica, Instituto Tecnol\'ogico de Aeron\'autica, CTA, 12228-900 S\~ao Jos\'e dos Campos, Brazil \\
                $^2$CFisUC, Department of Physics, University of Coimbra, P-3004 516 Coimbra, Portugal \\
                $^3$Instituto de Estudos Avan\c{c}ados and Instituto de F\'{\i}sica, Universidade de S\~ao Paulo,
                         Caixa Postal 66318, 05314-970 S\~ao Paulo, SP, Brazil \\
                $^4$Instituto de F\'{\i}sica Te\'orica (IFT), UNESP, S\~ao Paulo State University,
                        Rua Doutor Bento Teobaldo Ferraz 271, Bloco II, Barra Funda CEP 01140-070, S\~ao Paulo, Brazil}

\date{\today}

\begin{abstract}
In this paper, we propose a method to describe the many-body problem of electrons in honeycomb materials via the introduction of random fields which
are coupled to the electrons and have a Gaussian distribution. From a one-body approach to the problem, 
after integrating exactly the contribution of the random fields, one builds a non-hermitian and dissipative effective Hamiltonian with two-body interactions.
Our approach introduces besides the usual average over the electron field a second average over the random fields. The interplay of two averages
enables the definition of various types of Green's functions which allow the investigation of fluctuation-dissipation characteristics of the interactions that
are a manifestation of the many-body problem. In the current work we study only the dissipative term, through the perturbative 
analysis of the dynamics associated the effective Hamiltonian generated by two different kinds of couplings. For the cases analysed,
the eigenstates of the effective Hamiltonian are complex and, therefore, some of the states have a finite life time.
Moreover, we also investigate, in the mean field approximation, the most general parity conserving coupling to the random fields and compute the width of 
charge carriers $\Gamma$ as a function of the Fermi energy $E_F$. 
The theoretical prediction for $\Gamma (E_F)$ is compared to the available experimental data for graphene. 
The good agreement between $\Gamma_{theo}$ and $\Gamma_{exp}$ suggests that description of the many-body problem associated to the
electrons in honeycomb materials can indeed be done via the introduction of random fields.
\end{abstract}


\maketitle

\section{Introduction and Motivation}

In the dynamical description of a many-body problem it is common to approximate the full dynamics by one-body Hamiltonians. Within the one-body approach 
each constituent experiences the mean field due to all the other constituents particles. The original many-body system is replaced, in first approximation, by a 
system of independent particles whose wave function can be represented by a Slater determinant for fermion systems. Such an approximation does not provide 
an explanation for all the properties of the many-body problem and breaks down at some point. 
One can find in the literature various techniques to incorporate further dynamical details of the many-body dynamics in the description of such complex problems.

The nuclear shell model, the Hartree-Fock approximation used in the description of atomic and nuclear systems or the description of electrons in metals provide well known examples 
where one-body hamiltonians are used as starting point to analyse the full system. In what concerns the electronic properties of graphene,
typically and as a first approximation, one relies in a independent particle picture either by using a tight-binding model or the free Dirac equation.
The full problem, however, is a complex many-body problem even if one does not take into account impurities and other sources of 
disorder~\cite{Neto2009,Peres2010,Adam2009,Vozmediano2010}. 

From the technological point of view, the understanding on how the electronic properties of graphene change due to defects, impurities, sources of disorder or
deformations is crucial -- see e.g.~\cite{Araujo2012,Chaves2014} and references therein.
In graphene there are many possible sources of disorder such as the presence of charged impurities of the substrate, 
of resonant scatters, of structural defects, of strain fluctuations just to name a few. 
The different types of disorder can be translated into new electron interactions which, in principle, lead to new spectroscopic and transport observable phenomena.
Disorder introduces, typically, some random behaviour which requires, from the point of view of the theoretical approach, an averaging over various realisations of the
associated random background interaction.

From the point of view of the one-body problem, the sources of disorder can be viewed as local changes in the single site energy or changes in the 
distance and/or angles between the carbon $p_z$ orbitals. These can be translated into local effective interactions to be added to the original Hamiltonian. 
In graphene, the effects associated to a change of the overlap of the carbon orbitals can be rewritten introducing extra fields in the Dirac equation, 
see Sec. IV in~\cite{Neto2009}, or can be modelled using the Dirac equation in curved spaces, see e.g.~\cite{Chaves2014} and references therein. 

For the particular case of the Dirac equation in curved space, the effects of the deformation can be translated into a redefinition of the electron Fermi velocity $v_F$,
which becomes some kind of local field. That means that it is possible to define an effective Fermi velocity whose value is point dependent.
Another way of changing the Fermi velocity is doping the pristine material.
If the curvature of the graphene sheet together with disorder effects introduce random fluctuations, this effect can be studied by replacing $v_F$ 
by a random field whose mean value should reproduce the usual value $c/300$. 
Impurities or disorder are not the only mechanism  that can modify the value of the Fermi velocity. Indeed,
$v_F$ can differ from its standard value $c/300$ due to the many-body interactions~\cite{Popovici2013}. 
In this sense, fluctuations of the Fermi velocity can also be seen as a manifestation of the complex many-body problem in the dynamics of a single electron
in a graphene sheet. 

If the fluctuations of the Fermi velocity are a manifestation of the many-body problem, this suggests that the many-body problem itself can be mapped into a 
single particle equation via the introduction of random fields. In this case, an average over the random fields is required in order to produce meaningful results.
Recall that in Condensed Matter Physics, another way of taking into account various forces in the dynamics of the electron is to replace its mass by the effective mass
tensor which can be viewed also as another field to be taken into account. 

In the present work, we aim to explore the possibility of incorporating the dynamics of many-body problem through the introduction of random fields that
are coupled to the electron field. Although, herein we focus in graphene, the technique can be applied straightforwardly to any quantum system.

The use of stochastic and random process to understand a many-body problem has been explored, using a different perspective,
in  different areas of physics ranging from the study of the Dirac spectrum in QCD~\cite{Shuryak1993}, to nuclear reactions~\cite{Ericson1966,Verbaarschot1985}, 
complex systems~\cite{Dyson1962} or condensed matter~\cite{Lee1985,Beenakker2015} among others. 

In what concerns the use of random fields to investigate graphene properties, several examples can be found in the literature. 
In~\cite{Amanatidis2009} the electronic states for disordered graphene dots of finite size in relation to quantum chaotic behaviour were investigated.
In~\cite{Burgos2015}, the effects of disordered ripples on the conductivity of a monolayer graphene flake was studied.
Although invoking random fields or some sources of randomness, conceptually our proposal differs substantially from those mentioned.
Similar ideas can be applied to the dynamics of quantum mechanical systems coupled to white noise~\cite{Oliveira2015}.

As starting point to understand the electronic properties of graphene we use a continuum language. In this framework, the spinor field associated to the electrons
is a solution of the Dirac equation. The inclusion of further details of the many-body electronic problem is performed by introducing random scalar fields $\varphi$'s, 
which couple linearly to fermion operators. Moreover, one assumes that $\varphi$ have no dynamics but their values are distributed accordingly to a normal 
probability distribution. In principle, one could consider different types of probability distributions, but for a system with a large number of constituents, invoking the
central limit theorem, the Gaussian distribution is a natural choice. Furthermore, by employing a Gaussian distribution, the new average associated with
$\varphi$ can be performed exactly. Similarly to what was seen in Quantum Mechanics~\cite{Oliveira2015}, the integration over the random fields
leads to a new effective Hamiltonian $H_{eff}$ which contains only the original fermion fields and new two-body interactions, i.e. the one-body Hamiltonian defined
before introducing the coupling to the random field gives rise to a two-body Hamiltonian.
The new effective Hamiltonian is non-hermitian and to understand the dynamics associated to  $H_{eff}$ we look at different types of coupling.

The dynamical equation associated with $H_{eff}$ is formally equivalent to a nonlinear Schr\"odinger equation~\cite{Roberts2001}. This type of equation
is used in many branches of science as in hydrodynamical phenomena, nonlinear optics, nonlinear acoustics, quantum condensates, heat pulses in solids, 
quantum chaos and various dissipative phenomena. 

The fluctuations due to the random fields enable the definition of various types of correlation functions where the combined role of the fluctuations and
dissipation can be investigated simultaneously. Although, we write all the necessary formalism to analyse the contributions coming from the fluctuations to
any correlation function, herein we focus only on the dissipative part. We find that, for graphene, our theoretical approach enables to compute the charge carriers width
that is consistent with the recently measure results~\cite{Mak2014}.

The system discussed in the current work is targeted to graphene using the Dirac equation as a starting point. However, the approach is sufficiently general and 
can be applied to other problems. Natural candidates where the properties derived below are those materials which can also be described, in first approximation,
by a Dirac equation such as silicene or germanene.

The paper is organised as follows. In Sec.~\ref{Sec:Rand} we introduce the random fields and describe how one can integrate them exactly to
arrive into a quantum problem involving only the original fermion fields. In Sec.~\ref{Sec:fluct} we explore the interplay between the various
possible averages in connection to the dissipation-fluctuation points of view associated to a complex many-body problem.
In~\ref{Sec:exemplos}, we look at the dissipation aspects of the interaction and build perturbative solutions for two different types of couplings.
In~\ref{meanfield} a general parity conserving type of coupling to the random fields is considered and the quantum field problem is solved in
the Hartree approximation. We identify the corresponding charge carrier theoretical widths and compare our prediction with the recent available experimental
data. Finally, in Sec.~\ref{fim} we resume and conclude.

%==========================================================================
%==========================================================================
\section{Random Fields and Effective Actions \label{Sec:Rand}}

The continuum approach to the electronic properties of graphene relies on the massless Dirac equation and assumes a particle independent model.  
The Dirac equation for the electron field reads
\begin{equation}
   i \, \hbar \frac{\partial \psi}{\partial t} =  -i \, \hbar \, v_F \, \vec{\alpha} \cdot \nabla \psi \ ,
   \label{Eq:Dirac}
\end{equation}   
where $v_F$ is the Fermi velocity. For a system of units where $\hbar = 1$ and $v_F = 1$, the corresponding lagrangean density is given by
\begin{equation}
   \mathcal{L} = \overline\psi \, i \, \gamma^\mu \partial_\mu \, \psi \ .
   \label{Eq:L0}
\end{equation}   
Let us assume that $\psi$ is a quantum field and that the Fermi velocity fluctuates around its usual quoted value $v_F \approx c/300$. The fluctuations of $v_F$
can be attributed to disorder/defects of the pristine material and/or many-body interactions not taken into account in (\ref{Eq:Dirac}).
Let us introduce the dimensionless random field $\varphi$ to simulate this fluctuations and replace the lagrangean density (\ref{Eq:L0}) by
\begin{equation}
   \mathcal{L} = \overline\psi \, \left\{ i \, \gamma^0 \partial_0 + i \,  \left( 1 + \varphi \right) \vec{\gamma} \cdot \nabla \,  \right\}\psi \ .
   \label{Eq:L1}
\end{equation}   
If the system is described by the lagrangean density (\ref{Eq:L1}), the definition to the corresponding quantum field theory needs to build
the Green's generating functional $Z$ and, therefore, the average over the random field $\varphi$ has to be defined. 
For fluctuations of the Gaussian type it follows
\begin{eqnarray}
   Z_{eff} & = & \langle Z_\varphi \rangle \nonumber \\
   & = &
   \int \mathcal{D}\varphi ~ Z_\varphi ~  \exp\Bigg\{  - \frac{1}{2} \int d^3x \, d^3y \, \varphi (x) \, M(x,y) \, \varphi(y) \Bigg\} \nonumber \\
   &= &
   \int \mathcal{D}\varphi \, \mathcal{D}\overline\psi \, \mathcal{D}\psi 
   \exp\Bigg\{ i \int d^3x \, \bigg[ \overline\psi \, \left\{ i \, \gamma^0 \partial_0 + i \,  \left( 1 + \varphi \right) \vec{\gamma} \cdot \nabla \,  \right\}\psi \bigg] 
   - \frac{1}{2} \int d^3x \, d^3y \, \varphi (x) \, M(x,y) \, \varphi(y) \Bigg\}
\end{eqnarray}   
where $M$ is a symmetric invertible matrix with dimensions of mass to the power six. It turns out that the matrix $M$ has to satisfy further constraints in order
to arrive at a properly causal quantum field theory.
 
The effective action appearing in $Z_{eff}$,
\begin{equation}
    S[ \overline\psi , \psi , \varphi ] = \int d^3x \, \bigg[ \overline\psi \, \left\{ i \, \gamma^0 \partial_0 + i \,  \left( 1 + \varphi \right) \vec{\gamma} \cdot \nabla \,  \right\}\psi \bigg] 
   + \frac{i}{2} \int d^3x \, d^3y \, \varphi (x) \, M(x,y) \, \varphi(y)  \ ,
\end{equation}    
is nonlocal and has a pure imaginary term.
Given that the action is quadratic in $\varphi$, one can integrate over $\varphi$ and recover a generating functional which is a functional of the fermion degrees
of freedom only
\begin{equation}
   Z_{eff} \, = \, \int \mathcal{D}\overline\psi \, \mathcal{D}\psi 
   \exp\Bigg\{ i \int d^3x \, \bigg[ \overline\psi \,  i \, \gamma^\mu \partial_\mu \psi \bigg] 
+ \frac{1}{2} \int d^3x d^3y \, \left[ \overline\psi (x) \, \vec{\gamma} \cdot \nabla \psi (x) \right] \, M^{-1}(x,y) \, \left[ \overline\psi (y) \, \vec{\gamma} \cdot \nabla \psi (y) \right] \Bigg\} \ .
\label{Z:free}
\end{equation}   
The new effective fermionic action is given by
\begin{equation}
S_{eff} [ \overline\psi , \psi ]=
\int d^3x \, \bigg[ \overline\psi \,  i \, \gamma^\mu \partial_\mu \psi \bigg] 
- \frac{i}{2} \int d^3x d^3y \, \left[ \overline\psi (x) \, \vec{\gamma} \cdot \nabla \psi (x) \right] \, M^{-1}(x,y) \, \left[ \overline\psi (y) \, \vec{\gamma} \cdot \nabla \psi (y) \right]
\ .
\label{Eq:L_true}
\end{equation}
The integration over the fluctuations generates a non-local dissipative two-body interaction. The non-local interaction should not violate causality.
 A way out to avoid causality problems is to take $M$ proportional to a Dirac delta function. 
In this case,  $M^{-1}$ is also proportional to a Dirac delta function and the effective action (\ref{Eq:L_true}) is reduced to a local theory. 
Certainly, this is not the only possible solution to ensure that the theory (\ref{Eq:L_true}) is a causal theory.

The introduction of new random fields coupled to new fermionic operators generates new higher many-body effective interactions.
In general, fluctuations of an n-body operator gives rise to an effective 2n-body iteration.

The method can be extended to the case where the coupling of $\varphi$ to the quantum system is quadratic. 
The introduction of a quadratic coupling is equivalent to a redefinition of the matrix $M$. 
For example, for a quadratic coupling to the operator $A(x)$ given by
\begin{equation}
 \varphi^2 (x) A(x) \ ,
\end{equation} 
$M(x,y)$ should be replaced by
\begin{equation}
 M(x,y) - 2 \, i \, \frac{A(x) + A(y)}{2} \, \delta (x-y) \ .
\end{equation}
The price to pay being that the computation of $M^{-1}$ is not so simple but, at least formally, the integration over $\varphi$ can be performed. 
The new interactions have now terms of type $1/A(x)$, which possibly are not likely for a perturbative approach to the corresponding theory.

%%==========================================================
%%==========================================================
\section{Higher Order Statistical Fluctuations \label{Sec:fluct}}

In the previous section the integration over the random field $\varphi$ was discussed. This integration generated an effective interaction involving only the 
original fermion degrees of freedom and the corresponding Green's function generating functional $Z$ was built.
For any realisation of the random field $\varphi$ one can associate a generating functional
\begin{equation}
   Z_\varphi  [\overline\eta, \eta]\, = \, \int \mathcal{D}\overline\psi \, \mathcal{D}\psi 
   \exp\Bigg\{ i \, \int d^3x \, \mathcal{L} ( \varphi ; \overline\psi, \psi  )  + i \int d^3x \left[ \overline\psi (x) \eta(x) + \overline\eta (x) \psi (x) \right] \Bigg\} \ 
\end{equation} 
where $\eta$ and $\overline\eta$ are Grassmann sources that couple to $\overline\psi$ and $\psi$, respectively.
Recall that the coupling of the random field with the fermions is implicitly given in the lagrangean density. Following the procedure devised in Sec. \ref{Sec:Rand},
one can define
\begin{eqnarray}
   Z_{eff}[\overline\eta, \eta] & = & \langle  Z_\varphi  [\overline\eta, \eta] \rangle \nonumber \\
   & = & 
    \int \mathcal{D}\varphi ~ Z_\varphi [\overline\eta, \eta] ~\exp\bigg\{- \frac{1}{2} \int d^3x \, d^3y \, \varphi (x) \, M(x,y) \, \varphi(y) \bigg\}\nonumber \\
   & = &   
   \int \mathcal{D}\overline\psi \, \mathcal{D}\psi 
   \exp\Bigg\{ i \, S_{eff} [\overline\psi , \psi] +  i \int d^3x \left[\overline \psi (x) \eta(x) + \overline\eta (x) \psi (x) \right] \Bigg\} \ ,
\label{Z:freeJJ}
\end{eqnarray} 
where $S_{eff} [\overline\psi , \psi] $ depends on the coupling to $\varphi$ through $\mathcal{L}$.
In this way one builds a formal solution of the quantum field theory associated to a given Lagrangean density 
and taking into account the statistical average associated to $\varphi$.
Note that we are considering two different averages:
a quantum average associated to the functional integration over the fermion fields and a statistical average in connection with the random field.
The functional $Z_{eff}$ contains dissipative terms and from $S_{eff}$ one can identify a dissipative effective Hamiltonian. This particular case
will be explored in the next sections.

For the computation of a given correlation function, let us say a propagator $\langle 0 | \mathcal{T}\, \psi(x) \overline \psi (y) \, |0 \rangle$, depending on how
the quantum and statistical averages are performed, different types of Green's functions are accessed which can be identified with dissipation and fluctuations.
By integrating over the random fields, as considered
so far, one computes the propagator associated to the quantum field theory described by $Z_{eff}$:
\begin{equation}
   \langle 0 | \mathcal{T} \, \psi(x) \overline \psi (y) \, |0 \rangle_{eff}
    = \left. \frac{ \delta^2 Z_{eff}[\overline\eta , \eta]}{ \delta \overline\eta (x) \, \delta \eta (y)} \right|_{\overline\eta=\eta = 0}
    \,
    = \, \left.
    \frac{ \delta^2 \langle  Z_\varphi [\overline\eta , \eta] \rangle }{ \delta \overline\eta (x) \, \delta \eta (y)}  \right|_{\overline\eta=\eta = 0} 
    \,
    \neq
    \langle \left. \frac{ \delta^2 Z_\varphi[\overline\eta , \eta]}{ \delta \overline\eta (x) \, \delta \eta (y)}\rangle \right|_{\overline\eta=\eta = 0}  \ .
    \label{eq:propagadores}
\end{equation}   
In (\ref{eq:propagadores}), the last term requires the computation of the propagator for each realization of the random field,
i.e. to compute first the Green function
\begin{equation}
   \langle 0 | \mathcal{T} \, \psi(x) \overline \psi (y) \, |0 \rangle_\varphi
    = \left. \frac{ \delta^2 Z_\varphi [ \varphi, \overline\eta , \eta]}{ \delta \overline\eta (x) \, \delta \eta (y)} \right|_{\overline\eta=\eta = 0} \ ,
\end{equation}   
followed by the statistical average
\begin{equation}
\langle\langle 0 | \mathcal{T} \, \psi(x) \overline \psi (y) \, |0 \rangle\rangle
    \, = \, \int \mathcal{D}\varphi ~ \langle 0 | \mathcal{T} \, \psi(x) \overline \psi (y) \, |0 \rangle_\varphi ~
   \exp\Bigg\{ - \frac{1}{2} \int d^3x \, d^3y \, \varphi (x) \, M(x,y) \, \varphi(y) \Bigg\} \ .
\end{equation} 
This procedure defines a two point correlation function which carries the effects due to dissipation and fluctuations and from the definitions it follows that
\begin{equation}
\langle\langle 0 | \mathcal{T} \, \psi(x) \overline \psi (y) \, |0 \rangle\rangle
    \, = \, \langle 0 | \mathcal{T} \, \psi(x) \overline \psi (y) \, |0 \rangle_{eff} ~ + ~ \cdots,
    \label{Eq:correcoes}
\end{equation} 
where the corrections beyond the dissipative term $\langle 0 | \mathcal{T} \, \psi(x) \overline \psi (y) \, |0 \rangle_{eff}$ are associated with the
statistical fluctuations. The dissipation-fluctuations aspects of quantum mechanical systems with a large number of degrees of freedom expressed
by Eq. (\ref{Eq:correcoes}) can be found across different areas of physics, from nuclear and particle physics to quantum optics.

In order to substantiate the concepts under discussion, in the next sections we will explore the first term in (\ref{Eq:correcoes}), which is driven by dissipation,
for different types of couplings to the random fields.

%===========================================================================================
%===========================================================================================
\section{Dissipation from random fields: examples \label{Sec:exemplos}}

In order to study the dynamics associated to the coupling to the random fields, we will consider a particular simple realisation of the matrix $M$
which leads to a causal and local theory
\begin{equation}
   M(x,y) = i \, \frac{\delta ( x - y )}{\sigma^2} 
\qquad\mbox{ and, therefore, }\qquad
   M^{-1}(x,y) = -i \,\sigma^2\, \delta ( x - y ) \ ,
   \label{Eq:gauss}
\end{equation}
i.e. we are assuming that the random fields have a Gaussian distribution with width $\sigma$. Note that $\varphi$ is dimensionless and, therefore,
$\sigma^2$ has dimensions of $l^3$, where $l$ means length, or, for a system of units where $\hbar = v_F = 1$, has dimensions in $E^{-3}$, where $E$ stands
for energy.
In this case the effective lagrangean built from coupling
$\varphi$ to the Fermi velocity (\ref{Z:free}) reads
\begin{equation}
\mathcal{L}_{1} = 
\overline\psi \,  i \, \gamma^\mu \partial_\mu \psi ~ - ~ i \,
 \frac{\sigma^2}{2} \, \left[ \overline\psi \, \vec{\gamma} \cdot \nabla \psi \right]^2 \, .
 \label{L:1}
\end{equation}

As already discussed, the random field $\varphi$ can couple to any operator in the lagrangean. If graphene is on top of a given substrate, it develops a mass gap
and it is natural to add a mass term $-m_0 \overline\psi\psi$ to the free Dirac lagrangean. If the random field simulates corrections to the electron mass,
one expects that it should couple to a term that can be associated to an effective mass. In this case one can consider the following lagrangean density
\begin{equation}
\mathcal{L} =  \overline\psi \,  i \, \gamma^\mu \partial_\mu \psi ~ - ~ \left( m_0 + m_1 \varphi \right) \, \overline\psi \,  \psi 
\end{equation} 
prior to integration over $\varphi$. The integration over the random field gives rise  to the generating functional
\begin{equation}
   Z_{eff} \, = \, \int \mathcal{D}\varphi \, \mathcal{D}\overline\psi \, \mathcal{D}\psi 
   \exp\Bigg\{ i \int d^3x \, \bigg[ \overline\psi \,  i \, \gamma^\mu \partial_\mu \psi  - m \,\overline\psi \psi \bigg] 
- \frac{m^2_1}{2} \int d^3x d^3y \, \left[ \overline\psi (x) \,  \psi (x) \right] \, M^{-1}(x,y) \, \left[ \overline\psi (y) \,  \psi (y) \right] \Bigg\} \ 
\label{Z:mass}
\end{equation}   
and for a matrix $M$ as in (\ref{Eq:gauss}) it defines the effective lagrangean
\begin{equation}
\mathcal{L}_{2} = 
 \overline\psi \, \left(  i \, \gamma^\mu \partial_\mu - m_0 \right) \psi 
+ i \, \frac{m^2_1 \, \sigma^2}{2}  \left[ \overline\psi \,  \psi \right]^2 \ .
\label{L:2}
\end{equation}   
Note that for the cases considered so far, see Eqs. (\ref{L:1}) and (\ref{L:2}),
the new two-body interaction is proportional to the width of the Gaussian distribution $\sigma^2$ and, therefore, the original
lagrangean density is recovered for sufficiently small fluctuations, i.e. when $\sigma^2 \rightarrow 0$.

%===============================================================
%===============================================================
 \subsection{Wave Mechanics Approximation}
 
In order to illustrated the effects due to the new non-hermitian terms appearing in (\ref{L:1}) and (\ref{L:2}), we will  consider
the corresponding relativistic quantum mechanical problem in the small width approximation. The approach used in this section
ignores quantum fluctuations and reduce the quantum field theoretical problem to the computation of solutions of non-linear relativistic wave equations.

In order to find a solution of the non-linear Dirac equation we will use the representation considered in~\cite{Popovici2012} where
\begin{equation}
   \psi = \left( \begin{array}{c} \psi^b_K \\ \psi^a_K \\ \psi^a_{K'} \\ \psi^b_{K'} \end{array} \right) 
\end{equation}   
with the indices $a$, $b$ naming the sublattices $A$ and $B$, respectively, and $K$ and $K'$ referring to the Dirac points. 
For convenience, we introduce the two dimensional spinors
\begin{equation}
   u_K = \left( \begin{array}{c} \psi^b_K \\ \psi^a_K  \end{array} \right) 
   \qquad\mbox{ and }\qquad
   u_{K'} =  \left( \begin{array}{c} \psi^a_{K'} \\ \psi^b_{K'} \end{array} \right) 
\end{equation}
such that the four dimensional spinor can be written as
\begin{equation}
   \psi  = \left( \begin{array}{c} u_K \\ u_{K'} \end{array} \right)  \ .
\end{equation}
The Dirac matrices read
\begin{equation}
    \gamma^0 = \left( \begin{array}{cc} 0 & 1 \\ 1 & 0 \end{array} \right) \ ,
    \qquad\qquad
    \vec{\gamma} = \left( \begin{array}{cc} 0 & \vec{\sigma} \\ -\vec{\sigma} & 0 \end{array} \right)
    \qquad\qquad\mbox{ and }\qquad\qquad
    \gamma_5 = \left( \begin{array}{cc} -1 & 0 \\ 0 & 1 \end{array} \right)     \ ,
\end{equation}
where $\sigma^i$ represents a two dimensional Pauli matrix. Note that this representation differs from the usual chiral representation by a sign in $\gamma^0$.

%========================================================================
%========================================================================
\subsubsection{The massless case \label{massless}}

The wave equation associated with $\mathcal{L}_1$ reads
\begin{equation}
   i \frac{\partial \psi}{\partial t} = 
  \Bigg\{ - i \, \vec{\alpha} \cdot \nabla  + i \, \sigma^2 \Big[ \overline \psi \, \vec{\gamma}\cdot\nabla\psi \bigg] \, \vec{\alpha}\cdot\nabla \Bigg\} \psi \ ,
  \label{H1}
\end{equation}  
where $\vec{\alpha} = \gamma^0 \vec{\gamma}$. For $\sigma^2 = 0$ and setting
\begin{equation}
    \psi (x) = e^{i ( \vec{p} \cdot \vec{x} - E t)} \psi \ ,
\end{equation} 
where $ E = | \vec{p} |$, one can identify the following four independent solutions
\begin{equation}
  \psi^s_K = \left( \begin{array}{cc} \left[E - \vec{\sigma}\cdot\vec{p}\right] \chi^s \\ 0 \end{array} \right) 
  \qquad\qquad
  \psi^s_{K'} = \left( \begin{array}{cc} 0 \\ \left[E + \vec{\sigma}\cdot\vec{p}\right] \chi^s  \end{array} \right)   
\end{equation}
with
\begin{equation}
 \chi^+  = \left(\begin{array}{c} 1 \\ 0 \end{array}\right) \qquad\qquad\mbox{ and }\qquad\qquad  \chi^-   = \left(\begin{array}{c}0 \\ 1 \end{array}\right) \ .
\end{equation}
These solutions verify the  relations
\begin{eqnarray}
   \overline\psi^{s'}_K  \ \psi^s_K  =  \overline\psi^{s'}_{K'}  \ \psi^s_{K'} =  \overline\psi^{s'}_K  \ \psi^s_{K'}  = \overline\psi^{s'}_{K}  \ \psi^s_{K'} =  0 \qquad\quad
   \overline\psi^{s}_K  \, \gamma^0\,  \psi^s_K  =   \overline\psi^{s}_{K'}  \, \gamma^0\,  \psi^s_{K'} = ~ 2 \, E^2 \, N^2 
\end{eqnarray}   
where $N$ is a normalization factor. 

In (\ref{H1}) the term proportional to $\sigma^2$ can be treated as a perturbation to the free massless case. 
The two classes of solutions associated with $K$ and $K'$ are orthogonal to each other and the new interaction term in the hamiltonian
\begin{equation}
V_{new} =  i \, \sigma^2 \Big[ \overline \psi \, \vec{\gamma}\cdot\nabla\psi \bigg] \, \vec{\alpha}\cdot\nabla 
\end{equation}  
do not mix the two families of solutions, i.e. it does not break the intervalley symmetry. 
Therefore, one can consider the outcome of the perturbation separately on $\psi_K$ and $\psi_{K'}$. For the solutions associated to the Dirac point $K$
\begin{eqnarray}
   V^{s' \, s}_{K \, K} & = & 
   i \, \sigma^2  ~ \bigg[ \overline \psi^s_K \, \vec{\gamma}\cdot\nabla \psi^s_K \bigg] ~  \Bigg[ \left( \psi^{s'}_K \right)^\dagger \vec{\alpha}\cdot\nabla  \psi^s_K \Bigg] \nonumber \\
   & = & -\,  i \,E^2 \, \sigma^2 \,  \bigg[ \left( \psi^s_K \right)^\dagger \,  \psi^s_K \bigg] ~  \bigg[ \left( \psi^{s'}_K \right)^\dagger \,  \psi^s_K \bigg]  
\end{eqnarray}
and
\begin{equation}
   \left( V^{s' \, s}_{K \, K}  \right)  = 
  - \, i \, 4 \, N^4 \, E^5  \, \sigma^2 \, L^2 \,  ~  \left( \begin{array}{cc} E & - p_- \\ -p_+ & E \end{array} \right) 
\end{equation}
where $p_\pm = p_x \pm i p_y$. In perturbation theory for degenerate eigenstates, the first correction to the energy $\Delta$ is given by the solutions
$ \left| V  - \Delta \right| = 0 $ which are
\begin{equation}
\Delta = 0 \qquad\mbox{ or }\qquad \Delta = -  \, i \, 2 \, p^2 \, \sigma^2 \, n  \ ,
\label{Eq:DeltaVf}
\end{equation}
where $n$ should be interpreted as the carrier density. Choosing a normalisation of the Dirac spinor such that the integral of $\psi^\dagger \, \psi$ in a cubic box of size $L$ is 
unitary, then $n = 1/ L^2$.
Recall that $\sigma$ has dimensions of length to the power three and, therefore, $\Delta$ has dimensions of energy as expected. 
 
 It turns out that the new interaction coming from the integration over the Gaussian fluctuations removes the degeneracy 
 between the two sublattices A and B at the same point $K$. Of the two resulting states one remains massless and undamped, while the other has an energy given by
 $E = p - \, i \, 2 \, p^2 \, \sigma^2 \, n$ which describes a damped fermionic state. This results means that this damped state has a mean life
 $\tau =  ( 2 \, p^2 \, \sigma^2 \, n )^{-1}$  that is inversely proportional to the width of the Gaussian  distribution, to the momentum and to the charge carrier density.
The  damped state becomes stable, i.e. $\tau \rightarrow \infty$, when the Gaussian distribution becomes narrower and approach a delta distribution, for low momenta states
 and small carrier densities. The presence of a damped state tends to reduce the effective number of propagating modes that can contribute to transport properties. Note that
 for sufficiently small momentum and given that $\tau \propto p^2$, the ballistic transport properties of the charge carriers in graphene are protected against the noise 
 for this particular case of coupling.

So far we have considered only the solutions around the Dirac point $K$ but the results described also holds for the modes associated to the other Dirac point $K'$.

%=============================================================================
%=============================================================================
\subsubsection{The massive case \label{massive}}

Let us now discuss the case of the lagrangean density $\mathcal{L}_2$. The corresponding wave equations is given by
\begin{equation}
  \Bigg\{ i \, \gamma^\mu \partial_\mu  -  \bigg[ m_0  -  i \, m_1 \, \sigma^2 \, \overline \psi \psi \bigg] \, \Bigg\} \psi = 0 \  .
\end{equation}  
The interaction induced by the fluctuations contributes to the mass with a small negative pure imaginary term that is proportionally 
to the width of the gaussian noise $\sigma^2$ and also to the fermion condensate $\overline\psi \, \psi$. At the Hamiltonian level, the new term is 
\begin{equation}
   V_{new} = - i \, m^2_1 \, \sigma^2 \, \big[ \overline\psi \, \psi \big] \, \gamma^0 \ .
\end{equation}   

For a vanishing width, the solutions of the Dirac equation can be built in the usual way. Setting 
$\psi (x) = e^{-i p^\mu x_\mu} \, \psi$, where $p^0 = E = \sqrt{\vec{p}^{\, 2} + m^2_0}$, the positive energy solutions of the wave equation are
\begin{equation}
   \psi^s = \left( \begin{array}{r}  \chi^s \\  \frac{E + \vec{\sigma}\cdot\vec{p}}{m_0} \chi^s \end{array} \right) \ .
\end{equation}
Treating the new term $V_{new}$ using perturbation theory for degenerate states, a straightforward calculation gives
\begin{equation}
   \left( V^{s' \, s}_{K \, K}  \right)  = 
  - 4 \, i \, N^4 \, \sigma^2 \, E \, L^2 \,  ~  \left( \begin{array}{cc} E & p_- \\ p_+ & E \end{array} \right)  \ ,
\end{equation}
where $N^2$ is a normalisation factor. Then, the first order correction to the energy is given by
\begin{equation}
   \Delta = 4 \, i \, N^4 \, m^2_1 \, \sigma^2 \, E \, L^2 \, \left( - E \pm \sqrt{ E^2 - m^2_0 } \right) \, .
\end{equation}
If one chooses the same normalization as in the previous subsection, then $N^2 = m^2_0 / ( 2 E^2 L^2 )$ and
\begin{equation}
   \Delta =  i \, \sigma^2 \, \frac{m^2_1 m^2_0 \, n}{E^2 } \, \left( -1  \pm \sqrt{ 1 - \frac{m^2_0}{E^2} } \right) \, ,
\end{equation}
where $n$ should be read as the charge carrier density. It turns out that 
the first order correction to the energy is always a pure negative imaginary number and, therefore, the fluctuations associated to electron mass operator
give rise to unstable states and, therefore, the quantum states collapse for sufficiently large times. Note that width associated to the quantum states, i.e. the
imaginary part of the energy, is proportional to $\sigma^2$ and stability is recovered in the limit where $\sigma^2$ vanishes.

%%==========================================================
%%==========================================================
\section{Mean field approach to several random fields and charge carrier width \label{meanfield}}

\begin{figure}[t] %  figure placement: here, top, bottom, or page
   \centering
   \includegraphics[scale=0.4]{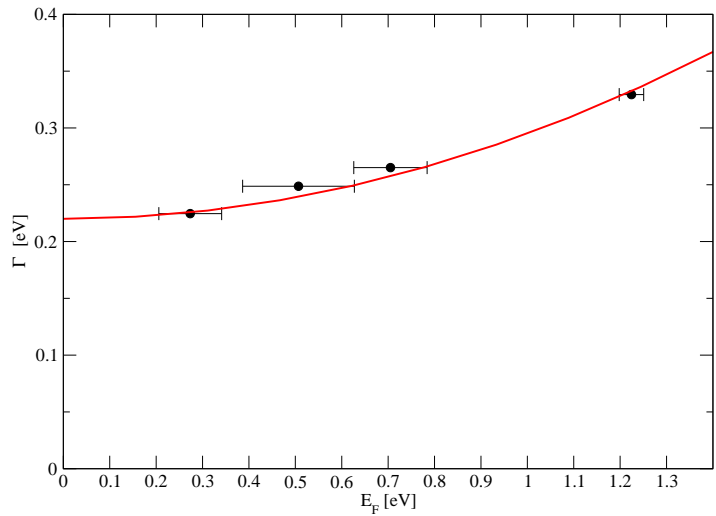} 
   \caption{The experimental widths as a function of the Fermi energy as measured in~\cite{Mak2014}. See text for details.}
   \label{fig:example}
\end{figure}

In Sec.~\ref{massless} and~\ref{massive} we considered a unique random field and studied the effect of considering a coupling to a single operator.
However, it is possible to consider various random fields, with each field coupled to a different operator. In order to simplify the
problem, herein we will focus on non-derivative terms, with the exception of that considered in Sec.~\ref{massless},
 and assume that parity is conserved. Then, for scalar random fields the most general Lagrangean density reads
\begin{equation}
   \mathcal{L}
    = \overline\psi \,  i \, \gamma^0 \partial_0  \, \psi  - m_0 \, \overline\psi \, \psi + i \, \left( 1 + g_\varphi \, \varphi \right) \overline\psi \,   \vec{\gamma} \cdot \nabla \,  \psi 
   +  g_s \, \xi_s \, \overline\psi \,  \psi
         +  g_0 \, \xi_0 \,\overline\psi \, \gamma^0 \psi
         + g_T \, \sum_{i,j} \frac{\xi_{ij}}{2}  \, \overline\psi \, \sigma^{ij} \psi \ ,
   \label{Eq:Lgeneral}
\end{equation}  
where $\varphi$, $\xi_s$, $\xi_0$ and $\xi_{ij} = - \xi_{ji}$ are Gaussian distributed random fields with widths given by $\sigma^2_\varphi$, $\sigma^2_s$,
$\sigma_0$ and $\sigma^2_{ij} = \sigma^2_{ji} $, respectively. Note that the coupling constants $g_s$, $g_0$ and $g_T$ have mass dimensions, 
while $g_\varphi$ is dimensionless. After the integration over the random variables one arrives at the following effective Lagrangean density
\begin{equation}
   \mathcal{L}_{eff} = \mathcal{L}_0  -
   \frac{i}{2} \, g^2_\varphi \,\sigma^2_\varphi  \left[ \overline\psi \,   \vec{\gamma} \cdot \nabla \,  \psi \right]^2
   + \frac{i}{2} \, g^2 _s \, \sigma^2_s\,\left[ \overline\psi \,  \psi \right]^2
    + \frac{i}{2} \,  g^2_0 \, \sigma^2_0 \, \left[ \overline\psi \, \gamma^0 \psi \right]^2
     + \frac{i}{4} \, g^2_T \, \sum_{i,j} \sigma^2_{ij} \, \left[ \overline\psi \, \sigma^{ij} \psi \right]^2 \ ,
   \label{Eq:Lgeneral2}
\end{equation}  
where
\begin{equation}
   \mathcal{L}_0  = \overline\psi \,  i \, \gamma^\mu \partial_\mu  \, \psi - m_0 \, \overline\psi \, \psi
\end{equation}  
is the usual free Dirac Lagrangean. The field equation associate to  (\ref{Eq:Lgeneral2}) is
\begin{equation}
  \bigg\{ i \, /  \!\!\! \partial   - m_0 - 
      i \, g^2_\varphi \,\sigma^2_\varphi  \left[ \overline\psi \,   \vec{\gamma} \cdot \nabla \,  \psi \right] \,   \vec{\gamma} \cdot \nabla
   +  i \, g^2 _s \, \sigma^2_s\,\left[ \overline\psi \,  \psi \right] 
    +  i \,  g^2_0 \, \sigma^2_0 \, \left[ \overline\psi \, \gamma^0 \psi \right] \,  \gamma^0 
     + \frac{i}{2} \, g^2_T \, \sum_{i,j} \sigma^2_{ij} \, \left[ \overline\psi \, \sigma^{ij} \psi \right] \,  \sigma^{ij} \psi \bigg\} \psi = 0 \ .
   \label{Eq:DiracNoisy}
\end{equation}
In the Hartree approximation the non-linear terms within the parenthesis are replaced by vacuum expectation values.
Setting $\psi (\vec{x},t) = e^{-i ( E t - \vec{p} \cdot \vec{x} )} \psi$ one can write to lowest order in perturbation theory
\begin{equation}
   \langle \overline\psi \, \vec{\gamma}\cdot\nabla \, \psi \rangle = i \bigg(  \langle E \overline\psi \, \gamma^0 \, \psi \rangle - m_0 \, \langle \overline\psi \, \psi \rangle \bigg)
    \ .
\end{equation}
Ignoring the contribution coming from the tensorial coupling, i.e. setting $g_T = 0$, equation (\ref{Eq:DiracNoisy}) can be written as
\begin{equation}
    \left(    \gamma^\mu \mathcal{P}_\mu - \mathcal{M} \right) \psi = 0 
\end{equation}
where
\begin{eqnarray}
   \mathcal{P}^0 & = & E + i \, g^2_0 \, \sigma^2_0 \, n \ , \\
   \mathcal{P}^j & = & p^j \bigg[ 1 - i \, g^2_\varphi \, \sigma^2_\varphi \bigg( \langle E \, \overline\psi \gamma^0\psi \rangle - m_0 \langle \overline\psi \, \psi \rangle \bigg) \bigg] 
       \ , \\
   \mathcal{M} & = & m_0 - i \, g^2_s \, \sigma^2_s \langle \overline\psi \, \psi \rangle \ 
\end{eqnarray}
and $n$ is the charge carrier density. Then, the energy-momentum dispersion relation for an electron in graphene is given by
\begin{eqnarray}
   E & = & E_0 - i \Gamma \nonumber \\
   E & \approx &\sqrt{ \vec{p}^{\, 2} + m^2_0} 
   - i \, \bigg\{ g^2_0 \, \sigma^2_0 \, n + m_0 \, g^2_s \, \sigma^2_s \frac{ \langle \overline\psi \, \psi \rangle }{\sqrt{ \vec{p}^{\, 2} + m^2_0}}
      + \frac{  \vec{p}^{\, 2} }{\sqrt{ \vec{p}^{\, 2} + m^2_0}} \, g^2_\varphi \sigma^2_\varphi \bigg( \langle E \, \overline\psi\gamma^0\psi\rangle - m_0 \langle \overline\psi\,\psi\rangle
      \bigg) \bigg\} \ .
\end{eqnarray}  
This results means that the many-body dynamics, here simulated by the coupling to the random fields, induces a width for the charge carriers given by
\begin{equation}
 \Gamma \approx g^2_0 \, \sigma^2_0 \, n + m_0 \, g^2_s \, \sigma^2_s \frac{ \langle \overline\psi \, \psi \rangle }{\sqrt{ \vec{p}^{\, 2} + m^2_0}}
      + \frac{  \vec{p}^{\, 2} }{\sqrt{ \vec{p}^{\, 2} + m^2_0}} \, g^2_\varphi \sigma^2_\varphi \bigg( \langle E \, \overline\psi\gamma^0\psi\rangle - m_0 \langle \overline\psi\,\psi\rangle
      \bigg) \ .
\end{equation}      
The vacuum expectation values are
\begin{equation}
   \langle E \, \overline\psi \gamma^0\psi \rangle = \frac{1}{4 \pi} \bigg[ \left( k^2_F + m^2_0 \right)^{3/2} - m^3_0 \bigg] \approx \frac{k^3_F}{4 \pi} \ ,
   \qquad   
   \langle \overline\psi \, \psi \rangle = \frac{m_0}{4 \pi} \bigg[ \sqrt{k^2_F + m^2_0} - m_0 \bigg] \approx \frac{m_0 k_F}{4 \pi}
       \qquad\mbox{and}\qquad
    n = \frac{k^2_F}{4 \pi} \ ,
\end{equation}   
where $k_F$ is the Fermi momentum and  $E_F \approx k_F$ is the Fermi energy. Then, one can write the electron width as
 \begin{equation}
 \Gamma \approx \frac{g^2_0 \, \sigma^2_0 \, E^2_F}{4 \pi} + 
 \frac{m^2_0 \, g^2_s \, \sigma^2_s \, E_F }{4 \pi \sqrt{ \vec{p}^{\, 2} + m^2_0}}
      + \frac{  \vec{p}^{\, 2} }{4 \pi \sqrt{ \vec{p}^{\, 2} + m^2_0}} \, g^2_\varphi \sigma^2_\varphi  \, E_F \bigg( E^2_F - m^2_0  \bigg)  \ .
\end{equation}      
For states close to the Fermi surface, the carrier width simplifes to
 \begin{equation}
 \Gamma \approx \frac{g^2_0 \, \sigma^2_0 \, E^2_F}{4 \pi} + 
 \frac{m^2_0 \, g^2_s \, \sigma^2_s  }{4 \pi }
      + \frac{ g^2_\varphi \sigma^2_\varphi  }{4 \pi }   \, E^2_F \bigg( E^2_F - m^2_0  \bigg) \ .
      \label{Eq:gamma}
\end{equation}      
This prediction of our theory can be compared against the experimental data of~\cite{Mak2014}, where the $\Gamma$ was measured for graphene.
It turns out that the experimental data $\Gamma$ as a function of the Fermi energy  is well reproduced by (\ref{Eq:gamma}) as can be seen
in Fig.~\ref{fig:example}.  An analysis of the data shows that the coefficient of $E^4_F$, associated to the fluctuations of the Fermi velocity, is much smaller
than all the others. Indeed, the data is well reproduced by $\Gamma  = 0.22 + 0.075 E^2_F$, where $\Gamma$ and $E_F$ are given in $eV$.
This suggests that the essential contribution to the
electron width in graphene are associated to the random fields that couple to the mass operator $\overline\psi \, \psi$ and the energy operator as given by
$\overline\psi\gamma^0\psi$. The contribution associated to the fluctuations of the Fermi velocity seem to give a marginal contribution to
$\Gamma$.

The prediction~(\ref{Eq:gamma}) gives a $\Gamma$ that is quadratic in the gap energy, i.e. in $m_0$. In this sense, changing the doping
or the graphene substrate dislocates the experimental data accordingly. 

In~\cite{Mak2014} besides the experimental measure of $\Gamma$, a theoretical calculation of the decay rates for highly excited electrons in graphene is made.
Although their theoretical computation underestimates the experimental value by 20-30\%, the authors look at the relative contributions to  $\Gamma$ due to
the electron-phonon and electron-electron interactions. It turns out that the contribution of the electron-phonon interaction seems to be independent
of the Fermi energy. On the other hand, the electron-electron interaction grows with $E_F$. 
It is curious to observe that in (\ref{Eq:gamma}) the constant term is associate to the electron mass term, 
while the term which grows with $E_F$ is connected with an interaction that is proportional to the energy of the many-electron system.

%%==========================================================
%%==========================================================
\section{Summary and Conclusions \label{fim}}

In this paper, we propose to study the many-body problem of electrons in graphene through the consideration of  random fields. Although, the
work described here uses graphene as inspiration, the conclusions of our investigation are valid to any material that can be described, in first
approximation, by a Dirac equation as e.g. silicene or germanene. The introduction of the random variables can be viewed as a way to incorporate statistically the details of the interaction in a many-body system.

Assuming that the random fields have a Gaussian distribution, it is possible to integrate their contribution exactly  using functional 
methods. The  interplay between the two types of averages considered, a functional integration over the electron fields and a
functional integration over the random fields, allows to investigate dissipation and fluctuations properties, which are a manifestation of the complex dynamics of the 
many-body problem.
Although, in the present work we provide all the ingredients to look at the fluctuation terms of the different Green's functions, we focused on the dissipative 
nature of the interaction.

The use of functional methods allows to integrate exactly the contribution of the random fields and identify a non-hermitian and dissipative effective Hamiltonian
$H_{eff}$. Starting from a one-body Hamiltonian, the integration of the random fields builds an Hamiltonian with two-body interactions. The non-hermitian and dissipative
terms in $H_{eff}$ are proportional to the fluctuations of the random fields. Furthermore, they vanish when the  distribution associated to the random fields
approaches a delta function distribution, i.e. in the limit of zero fluctuations.

The analysis of the perturbative solution for two different types of coupling shows that the eigenstates of $H_{eff}$ have, typically, complex eigenvalues and, therefore,
the corresponding states have a finite lifetime. For the so called massless case, see Sec.~\ref{massless}, one of the electron states is stable but the other
gets a width proportional to the charge carrier momentum. From small charge carrier momentum, the width essentially vanishes, and the corresponding
quantum states are effectively stable.
On the other hand, the states of $H_{eff}$ coming from coupling the random fields to a mass term, see Sec.~\ref{massive},  produces eigenstates
that have always a finite width and lifetime. The perturbative solution of the cases devised in Secs~\ref{massless} and~\ref{massive} gives a first flavour to
the full quantum field theoretical problem associated to $H_{eff}$.

In Sec.~\ref{meanfield}, we investigated a more general type of coupling between the electron and the random fields and attempted a Hartree mean field solution of the many-body problem with the effective Hamiltonian. The computation of the $H_{eff}$ eigenvalues shows that the charge carriers have, in general, finite widths. Furthermore,
we look at how the width $\Gamma$ changes with the charge carrier Fermi energy $E_F$. 
This allows us to compare our prediction
for $\Gamma (E_F)$ with recent available experimental data for graphene. 
The good agreement between the theoretical prediction for $\Gamma (E_F)$ and the experimental data suggests that, indeed,
one can describe the many-body problem of an electron in graphene, or other similar materials,
via the introduction of random fields as considered in the current work. 

A final remark concerning the effect of fluctuations on the averaged Green's function. The procedure described above using functional integration and ending up dealing with an 
effective theory governed by a non-Hermitian Hamiltonian, can also be used to analyze the fluctuation contribution. The fluctuation effects are embedded in the equation that defines 
the full Green's function, which is described by the so-called Pastur equation \cite{DMW07}. The analytical solution of this equation is difficult to obtain. On the other hand, the 
Pastur equation is amenable to numerical treatment, which will be investigated in a future publication.

%%==========================================================
%%==========================================================
\section*{Acknowledgements}

The authors acknowledge financial support from the Brazilian
agencies FAPESP (Funda\c c\~ao de Amparo \`a Pesquisa do Estado de
S\~ao Paulo) and CNPq (Conselho Nacional de Desenvolvimento
Cient\'ifico e Tecnol\'ogico). OO acknowledges financial support from grant 2014/08388-0 from S\~ao Paulo Research Foundation (FAPESP). MSH acknowledges  a CAPES/ITA
PVS Fellowship, CEPID/FAPESP, and INCT/CNPq.

%=======================================================================
%=======================================================================

\end{document}